\documentclass[conference]{IEEEtran}

\usepackage{amsmath,amssymb,amsfonts}
\usepackage{amsthm}
\usepackage{graphicx}
\usepackage{cite}
\usepackage{algorithm}
\usepackage{algorithmic}
\usepackage{booktabs}
\usepackage{url}
\usepackage{xcolor}

\DeclareMathOperator{\atan2}{atan2}

\newtheorem{remark}{Remark}

\newtheorem{propositi. on}{Proposition}

\theoremstyle{definition}

\newcommand{\Sone}{\mathbb{S}^{1}}
\newcommand{\wrap}{\mathrm{wrap}}
\newcommand{\LPF}{\mathrm{LPF}}

\begin{document}

\title{LO-Free Receiver: Next-Gen Low-Power Joint Communication and Sensing}

\author{
     Hasan Atalay Günel\IEEEauthorrefmark{1}\IEEEauthorrefmark{3}, Mohaned Chraiti\IEEEauthorrefmark{2}, and Ali Görçin\IEEEauthorrefmark{1}\IEEEauthorrefmark{3} \\
    \IEEEauthorrefmark{1}Communications and Signal Processing Research (HİSAR) Lab., T{\"{U}}B{\.{I}}TAK B{\.{I}}LGEM, Kocaeli, Turkey \\
 \IEEEauthorrefmark{2} Department of Electronics Engineering, Sabanci University, Istanbul, Turkey\\ 
\IEEEauthorrefmark{3}
Department of Electrical and Electronic Engineering, Istanbul Technical University, Istanbul, Turkey.\\
    Emails:
    hasan.gunel@tubitak.gov.tr,  
    mohaned.chraiti@sabanciuniv.edu,
    and 
    ali.gorcin@tubitak.gov.tr.
}

\maketitle

\begin{abstract}
This paper introduces and analyzes Spatial Phase Manifold Communications (SPMC), a paradigm that facilitates joint communication and sensing (JCAS) over Local Oscillator (LO) free receiver. Information is embedded in, and recovered from, the relative spatial phase between antennas. In contrast to conventional coherent receivers that rely on LOs and on channel estimation/equalization, SPMC exploits antenna-domain correlation to form a baseband observable that is a function of inter-antenna phase differences. Since these phase differences are fundamentally tied to Direction-of-Arrival (DoA) and vice-versa, the formulation recasts communication and sensing as inference over the unit-circle manifold and thus naturally supports JCAS decomposition, i.e., data and spatial sensing are encoded and recovered through DoA signatures. We develop a comprehensive framework comprising: (i) a manifold-domain signal model and corresponding phase-alphabet design; (ii) an LO-free quadrature spatial-correlator receiver architecture that resolves the phase-sign ambiguity without requiring an LO; and (iii) an analysis of error probability and sensing precision, including robustness to phase noise. The proposed paradigm is particularly suited to massive Internet-of-Things (IoT) deployments, for which hardware simplicity, LO distribution cost, power consumption, and seamless sensing integration are critical, especially at millimeter-wave and higher carrier frequencies.
\end{abstract}


\section{Introduction}

The evolution of the Internet of Things (IoT) toward a fully connected world comes with device-level constraints on power consumption, cost, and hardware complexity~\cite{IoT2015al,lowrf2018gao}. These constraints become more challenging to meet at millimeter-wave (mmWave) and higher frequency bands, where factors such as elevated hardware cost, increased power consumption, and radio-frequency impairments—most notably phase noise—impose additional performance and design burdens on receiver front-end architectures \cite{yang2019hard}. In parallel, emerging use cases increasingly require advanced features such as joint communication and sensing (JCAS) capabilities at the endpoint \cite{he2024isac}. Under these conditions, conventional coherent receivers—which rely on local oscillators (LOs) and carrier/phase tracking—can be costly in power and complexity and can be sensitive to oscillator impairments. These considerations motivate alternative receiver architectures that recover the desired information with reduced complexity.

\subsection{Related Work}

Minimizing LO power consumption has been a primary objective in low-power radio design, particularly for IoT receivers. Prior art has explored LO-relaxed and LO-free front-ends that overcome the need for continuous LO operation by leveraging envelope detection, square-law nonlinearities, or self-mixing to downconvert Radio Frequency (RF) signals for digitization~\cite{wentzloff2020review,wurx2022,jlpea2025wvrx,selfdemod2010}. By construction, these architectures typically operate on noncoherent magnitude- or beat-based observables. This is well-suited for low-power detection, but it can suppress the absolute and, critically, the inter-antenna relative phase structure that underpins sensing functions (e.g., Direction-of-Arrival (DoA) inference and beamspace processing)~\cite{godara97}. As a result, sensing functionality is often not a native byproduct of the recovered observable and instead requires additional circuitry, reference signals, or dedicated resources~\cite{zhang2022}.

Another research line considers Simultaneous Wireless Information and Power Transfer (SWIPT) receiver architectures in which information decoding is integrated into the energy-harvesting path (integrated rectifier--receiver), leveraging rectifier nonlinearity to enable LO-free or LO-relaxed detection~\cite{rajabi2018irr}. While such designs can reduce receiver-side LO requirements, their decision variables are derived from rectification-generated baseband components and therefore do not directly provide an observable that preserves inter-antenna relative phase~\cite{godara97}. Consequently, exploiting JCAS may require additional front-end structure or processing beyond the rectifier output.

\subsection{Problem statement}
The feasibility of achieving JCAS under fully LO-free reception remains unresolved. In particular, the noncoherent observables conventionally employed for LO-free operation typically discard phase information, whereas DoA sensing fundamentally relies on relative phase differences across the antenna array. Consequently, it is still not clear how to construct a baseband statistic that simultaneously (i) preserves antenna-domain phase differentials in the absence of LO-based carrier and phase synchronization, and (ii) supports both reliable data detection and accurate spatial inference from the same set of received measurements.

\subsection{Contributions}

We introduce Spatial Phase Manifold Communications (SPMC), a unified signaling and receiver framework designed to enable LO-free JCAS. In the SPMC architecture, we consider a receiver equipped with two or more antennas that forms antenna-domain correlations. The transmit antennas are assumed to form a sparse configuration, i.e., a Distributed Antenna System (DAS), in which the active radiating element is alternated according to the information sequence. Consequently, each transmitted symbol induces a distinct DoA signature at the receiver. The latter exploits these DoA-dependent spatial signatures for detection, while the same DoA estimates naturally support sensing. A key advantage is that the receiver can operate without coherent downconversion, by relying on correlation across antenna ports rather than an LO-based I/Q chain. Our main contributions are summarized as follows:

\begin{itemize}
  \item We introduce a DoA-signature-based JCAS in which information is conveyed through controlled switching across spatially sparse transmit antennas, yielding distinct angle-dependent signatures at a multi-antenna receiver.
  \item We develop an LO-free receive front-end based on antenna-domain correlation, enabling detection from low-frequency correlation observables rather than coherent I/Q demodulation.
  \item We characterize design/sufficient conditions under which the resulting spatial signatures are separable for reliable detection.
  \item We analyze performance under practical impairments, including fading and Automatic Gain Control (AGC) uncertainty, and quantify the resulting sensitivity/robustness.
  \item We benchmark against coherent receivers in terms of error rate and DoA estimation.
\end{itemize}

The remainder of the paper is organized as follows. Section~\ref{sec:system_model} presents the system geometry and the proposed LO-free correlation-based receiver. Section~\ref{sec:inference} develops the manifold-domain inference framework, including the Maximum Likelihood (ML) detection for communication and Cram\'er--Rao Lower Bounds (CRLB) based sensing limits. Section~\ref{sec:sim} provides numerical evaluation and benchmarking, and Section~\ref{sec:conclusion} concludes the paper.

\section{Spatial Phase Manifold Communications}\label{sec:system_model}
\subsection{System Model}

\begin{figure*}
    \centering
    \includegraphics[width=0.7\linewidth]{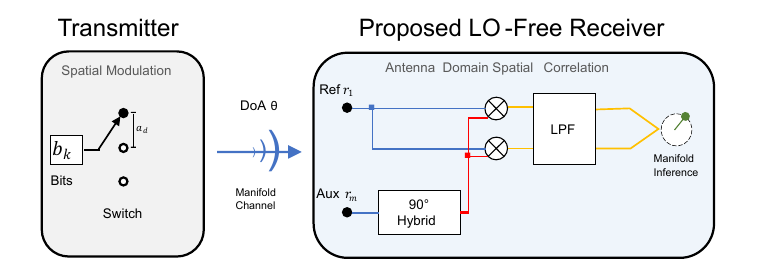}
        \caption{SPMC transmitter and receiver design.}\label{fig:weighted_comparison}
    \vspace{-0.3cm}
\end{figure*}

We consider a DAS transmitter with inter-element spacing $a_{d}$, in which transmit antennas are placed far enough apart that activating different antennas yields angularly resolvable dominant directions of arrival at the receiver array. The information sequence is thus embedded in the pattern of active-antenna switching, which induces a discrete DoA sequence $\{\phi_k\}_k$ at the receiver. This DoA sequence acts as the physical-layer signature that conveys the data, and it can also be leveraged for spatial sensing. An illustrative diagram of the transmitter and receiver is provided in Fig.~\ref{fig:weighted_comparison}.

The receiver employs an $M$-antenna array. For concreteness and without loss of generality for the subsequent correlation receiver, we consider a uniform linear array (ULA) with receiver inter-element spacing $d$ along a baseline. Let $\phi_k$ denote the DoA of the dominant component at symbol time $k$.\footnote{mmWave and higher-frequency links often exhibit sparse multipath channels. It is common that a strongest component (line-of-sight (LOS) or quasi-LOS) dominates the spatial signature, while residual multipath (e.g., a small number of weak specular reflections) can be modeled as an effective phase perturbation \cite{akdeniz2014mmwave}.} For an ULA, the per-element spatial phase increment induced by $\phi_k$ is
\begin{equation}
\Delta\theta_{\text{DoA}}(k) \triangleq \frac{2\pi d}{\lambda}\sin(\phi_k),
\label{eq:doa_phase}
\end{equation}
where $\lambda$ is the carrier wavelength. The resulting array-manifold phase difference across antennas satisfies
\begin{equation}
\alpha_m(k) - \alpha_1(k) = (m-1)\,\Delta\theta_{\text{DoA}}(k), \qquad m=1,2,\ldots,M,
\label{eq:phase_progression}
\end{equation}
where $\alpha_m(k)$ encodes the spatial phase at time $k$. The inter-antenna phase difference scales linearly with the baseline length. The receiver does not attempt to track the absolute carrier phase; instead, it infers the spatial phase differences (equivalently, the manifold parameter) corrupted by an effective phase disturbance: 

\begin{equation}
\wrap\!\left(\Delta\theta_{\text{DoA}}(k) + \nu_k\right),
\label{eq:phase_superposition}
\end{equation}
where $\nu_k$ aggregates residual multipath, receiver noise after correlation/normalization, and model mismatch, and $\wrap(\cdot)$ maps to $(-\pi,\pi]$. The RF waveform received at the $m$-th antenna is modeled as
\begin{equation}
r_m(t) = A_m \cos\!\big(2\pi f_c t + \psi(t) + \alpha_m(k)\big) + w_m(t),
\label{eq:rf_model_M}
\end{equation}
 where $A_m$ captures channel amplitude and front-end gain effects, $\psi(t)$ is an unknown phase noise at the transmitter, and $w_m(t)$ denotes bandpass receiver noise. The key structural property is that $\psi(t)$ is common across all the receiver antennas and can be eliminated by antenna-domain correlation.

\subsection{Receiver Architecture: Antenna-Domain Correlation}
The receiver forms antenna-domain correlations directly at RF and avoids coherent downconversion. For any antenna pair $(1,m)$, we compute the product
\begin{equation}
y_m(t) = r_1(t)\,r_m(t),
\end{equation}
followed by low-pass filtering $z_{c,m}(t) = \LPF\{y_m(t)\}$. Neglecting noise cross-terms and high-frequency components, the dominant low-frequency component is
\begin{equation}
z_{c,m}(t) \approx \frac{A_1A_m}{2}\cos\!\big(\Delta\theta_{1m}(t)\big) + n_{c,m}(t),
\label{eq:cos_observable_M}
\end{equation}
where $\Delta\theta_{1m}(t)\triangleq \alpha_m(k)-\alpha_1(k)$ over the $k$-th symbol duration and $n_{c,m}(t)$ is an effective low-frequency noise term. Importantly, the phase noise $\psi(t)$ cancels out as shown in \eqref{eq:cos_observable_M}. Hence, these observables can be formed without an LO.

The cosine-only observable does not capture the sign ambiguity when mapping $z_{c,m}$ to $\Delta\theta_{1m}$. To resolve this without an LO, a quadrature correlation branch can be implemented via an RF $90^\circ$ hybrid (or an equivalent quadrature network) that generates a $\pi/2$-shifted version of one input prior to correlation. Let
\begin{equation}
\tilde{r}_m(t) \approx A_m \cos\!\big(2\pi f_c t + \psi(t) + \alpha_m(k) + \tfrac{\pi}{2}\big).
\end{equation}
Then define
\begin{equation}
\begin{split}
    z_{s,m}(t) &= \text{LPF}\left\{r_1(t)\,\tilde{r}_m(t)\right\} \\
    &\approx \frac{A_1A_m}{2}\sin\!\big(\Delta\theta_{1m}(t)\big) + n_{s,m}(t),
\end{split}
\label{eq:sin_observable_M}
\end{equation}
with $n_{s,m}(t)$ denoting the effective quadrature-branch noise.

Collecting the in-phase and quadrature correlation outputs yields, for each baseline $(1,m)$, a two-dimensional manifold observation:
\begin{equation}
\mathbf{z}_m(t)=
\begin{bmatrix}
z_{c,m}(t)\\ z_{s,m}(t)
\end{bmatrix}
\approx
\frac{A_1A_m}{2}
\begin{bmatrix}
\cos\!\big(\Delta\theta_{1m}(t)\big)\\ \sin\!\big(\Delta\theta_{1m}(t)\big)
\end{bmatrix}
+
\mathbf{n}_m(t),
\label{eq:zs_zc_vector_M}
\end{equation}
which provides an unambiguous representation (up to wrapping) of the spatial phase difference on $\Sone$.

Amplitude terms $A_1A_m$ may fluctuate due to fading and gain/AGC effects. We therefore normalize each baseline output as
\begin{equation}
\hat{\mathbf{u}}_m(t) =
\frac{\mathbf{z}_m(t)}{\|\mathbf{z}_m(t)\|+\epsilon}
\approx
\begin{bmatrix}
\cos\!\big(\Delta\theta_{1m}(t)\big)\\ \sin\!\big(\Delta\theta_{1m}(t)\big)
\end{bmatrix}+\frac{\mathbf{n}_m(t)}{\|\mathbf{z}_m(t)\|+\epsilon},
\label{eq:normalize_M}
\end{equation}
with a small $\epsilon>0$ for numerical stability. In an $M$-antenna array, the set $\{\hat{\mathbf{u}}_m(t)\}_{m=2}^M$ furnishes multiple noisy samples of the same underlying DoA-induced phase progression \eqref{eq:phase_progression}, enabling robust inference (e.g., via baseline fusion) while remaining insensitive to unknown amplitude scaling and without requiring coherent down-conversion.

\section{Manifold-Domain Inference: Detection and Sensing}
\label{sec:inference}
In this section, we develop statistically grounded multi-baseline fusion, a ML detection of the active DAS antennas, and a DoA estimation with analytical localization characterizations.


\subsection{ML Fusion Across Baselines}

Let $k$ index the signaling interval. Each baseline $(1,m)$ can for a symbol-rate sample by averaging the correlator outputs over the $k$-th interval and then normalizing as in \eqref{eq:normalize_M}. Denote the resulting two-dimensional sample by $\hat{\mathbf{u}}_{m,k}\in\mathbb{R}^2$, and define its complex representation
\begin{equation}
q_{m,k} \triangleq \hat u_{c,m}(k) + j\,\hat u_{s,m}(k),
\label{eq:qmk_def}
\end{equation}
where $\hat u_{c,m}(k)$ and $\hat u_{s,m}(k)$ are the in-phase and quadrature components of $\hat{\mathbf{u}}_{m,k}$. The noiseless manifold point for baseline $(1,m)$ is
$e^{j\Delta\theta_{1m}(k)}$ with $\Delta\theta_{1m}(k)=(m-1)\Delta\theta_{\text{DoA}}(k)$ from \eqref{eq:phase_progression}. We model residual impairments after correlation/normalization via a circular perturbation,
\begin{equation}
q_{m,k} = e^{j\Delta\theta_{1m}(k)}\,e^{j\varepsilon_{m,k}},
\label{eq:qmk_noise}
\end{equation}
where $\varepsilon_{m,k}$ is a wrapped noise capturing residual multipath, correlator noise, and hardware mismatch. A convenient and widely used likelihood model on the unit circle is the Von Mises (also known as Tikhonov) distribution \cite{fisher1993circular}
\begin{equation}
p\!\left(q_{m,k}\,\big|\,\Delta\theta_{\text{DoA}}(k)\right)
\propto
\exp\!\Big(
\kappa_m \Re\{ q_{m,k} e^{-j(m-1)\Delta\theta_{\text{DoA}}(k)}\}
\Big),
\label{eq:vonmises_like}
\end{equation}
where $\kappa_m\ge 0$ is a concentration (reliability) parameter that can absorb signal-to-noise ratio (SNR) differences across baselines. The model in \eqref{eq:vonmises_like} is also consistent with the small-noise regime of wrapped Gaussian perturbations on $\mathbb{S}^1$ \cite{fisher1993circular} and leads to tractable estimators/detectors with clear signal-processing interpretation.

Given $\{q_{m,k}\}_{m=2}^M$, the log-likelihood induced by \eqref{eq:vonmises_like} is
\begin{equation}
\ell\!\left(\Delta\theta\right)
=
\sum_{m=2}^{M} \kappa_m \Re\!\Big\{ q_{m,k} e^{-j(m-1)\Delta\theta}\Big\}
+\text{const}.
\label{eq:ll_delta}
\end{equation}
Hence, the ML estimate of the per-element phase increment $\Delta\theta_{\text{DoA}}(k)$ is the one-dimensional maximization
\begin{equation}
\Delta\widehat{\theta}_{\text{DoA}}(k)
=
\arg\max_{\Delta\theta\in(-\pi,\pi]}
\sum_{m=2}^{M} \kappa_m \Re\!\Big\{ q_{m,k} e^{-j(m-1)\Delta\theta}\Big\}.
\label{eq:ML_delta}
\end{equation}
The criterion in \eqref{eq:ML_delta} is a matched filter on the array manifold: it aligns the measured phasors with the hypothesized phase progression $\{e^{j(m-1)\Delta\theta}\}$.

\begin{remark}
The fusion rule is practically stable in the sense that it avoids an explicit phase-extraction and phase-unwrapping step across baselines. Methods that first compute  the angle $\angle(q_{m,k})$ for each antenna pair and then ``unwrap'' these phases can suffer from occasional $2\pi$ jumps when a baseline has low SNR, which may cause large, discontinuous errors. In contrast, the defined objective operates directly on the unit-circle observations through $\Re\{\cdot\}$ and remains robust even when some antenna-pair measurements are noisy; unreliable baselines contribute weakly (or can be down-weighted) rather than destabilizing the estimate.
\end{remark}

At high SNR, a closed-form approximation is tractable. In fact, when $\varepsilon_{m,k}$ is small, $\angle(q_{m,k})$ can be approximated locally by $ (m-1)\Delta\theta_{\text{DoA}}(k)+\varepsilon_{m,k}$. In this regime, \eqref{eq:ML_delta} can be well-approximated by a weighted least-squares fit in the phase domain \cite{kay1993estimation,vanTrees2001detection}:
\begin{equation}
\widehat{\Delta\theta}_{\text{DoA}}(k)
\approx
\frac{\sum_{m=2}^M \kappa_m (m-1)\,\angle(q_{m,k})}{\sum_{m=2}^M \kappa_m (m-1)^2},
\label{eq:WLS_delta}
\end{equation}
followed by wrapping to $(-\pi,\pi]$. This approximation makes explicit the baseline-squared sensitivity: longer baselines provide higher phase slope with respect to $\Delta\theta_{\text{DoA}}(k)$.


\subsection{Communication: Error probability}

Let the DAS comprise $N_{\mathrm{tx}}$ spatially sparse antennas. Under the sparse angular regime, activating the $i$th antenna induces an (approximately) deterministic dominant DoA $\phi^{(i)}$ at the receiver, and hence a corresponding per-element phase increment
\begin{equation}
\Delta\theta^{(i)}_{\text{DoA}} \triangleq \frac{2\pi d}{\lambda}\sin\!\big(\phi^{(i)}\big).
\label{eq:delta_i}
\end{equation}
At time $k$, the active index $I_k\in\{1,\ldots,N_{\mathrm{tx}}\}$ defines the hypothesis
$\mathcal{H}_i: \Delta\theta_{\text{DoA}}(k)=\Delta\theta^{(i)}_{\text{DoA}} $.
From \eqref{eq:vonmises_like}, the ML detector is
\begin{equation}
\widehat{I}_k
=
\arg\max_{i\in\{1,\ldots,N_{\mathrm{tx}}\}}
\sum_{m=2}^{M} \kappa_m \Re\!\Big\{ q_{m,k} e^{-j(m-1)\Delta\theta^{(i)}_{\text{DoA}} }\Big\}.
\label{eq:ML_detector}
\end{equation}
Consider two indices $i$ and $j$ with wrapped separation
$\Delta_{ij}\triangleq \wrap(\Delta\theta^{(i)}_{\text{DoA}} -\Delta\theta^{(j)}_{\text{DoA}} )$.
Under the small-noise phase-domain approximation $\angle(q_{m,k})\approx (m-1)\Delta\theta^{(i)}_{\text{DoA}} +\varepsilon_{m,k}$ with i.i.d.\ $\varepsilon_{m,k}\sim \mathcal{N}(0,\sigma_\varepsilon^2)$, the pairwise error probability of the ML test admits the approximation
\begin{equation}
\mathbb{P}(i\to j)
\approx
Q\!\left(
\frac{|\Delta_{ij}|}{2\sigma_\varepsilon}\sqrt{\sum_{m=2}^M \kappa_m (m-1)^2}
\right),
\label{eq:pep}
\end{equation}
highlighting two fundamental design levers: (i) the minimum wrapped separation among $\{\Delta\theta^{(i)}_{\text{DoA}} \}$ induced by the DAS geometry, and (ii) the effective aperture term $\sum_{m=2}^M (m-1)^2$ provided by multi-antenna fusion.


\subsection{Sensing: Fisher-Information and CRLB Scaling}

We next quantify the intrinsic DoA accuracy implied by the manifold samples. Recall the baseline index $m\in\{2,\ldots,M\}$ and the per-element phase increment parameter
$\Delta\theta_{\text{DoA}}(k)$ defined in \eqref{eq:doa_phase}--\eqref{eq:phase_progression}. For notational brevity in this subsection, fix a time index $k$ and write $\Delta\theta \equiv \Delta\theta_{\text{DoA}}(k)$.

For each baseline $m\in\{2,\ldots,M\}$, we have $\Delta\theta_{1m}(k)\triangleq \angle(q_{m,k})\in(-\pi,\pi]$. Under the von Mises likelihood \eqref{eq:vonmises_like}, the conditional density of $\Delta\theta_{1m}(k)$ given $\Delta\theta$ is

\begin{equation}
\begin{split}
&p(\Delta\theta_{1m}(k)\mid \Delta\theta)
= \frac{1}{2\pi I_0(\kappa_m)} \\
&\qquad \times \exp\!\Big(\kappa_m \cos\!\big(\Delta\theta_{1m}(k)-(m-1)\Delta\theta\big)\Big),
\end{split}
\label{eq:vonmises_phi}
\end{equation}
where $I_0(\cdot)$ is the modified Bessel function of the first kind (order zero). The corresponding log-likelihood is
\begin{equation}
\begin{split}
&\ell_m(\Delta\theta)\\
&\triangleq
\log p(\Delta\theta_{1m}(k)\mid \Delta\theta) \\
&=
-\log\!\big(2\pi I_0(\kappa_m)\big)
+\kappa_m \cos\!\big(\Delta\theta_{1m}(k)-(m-1)\Delta\theta\big).
\end{split}
\label{eq:ll_m}
\end{equation}
Since the normalizing term does not depend on $\Delta\theta$, derivatives are governed by the cosine term. The score (first derivative) is
\begin{equation}
\begin{split}
\frac{\partial \ell_m(\Delta\theta)}{\partial \Delta\theta}
&= \kappa_m \frac{\partial}{\partial \Delta\theta} \cos\!\big(\Delta\theta_{1m}(k)-(m-1)\Delta\theta\big)
\\
&= \kappa_m (m-1) \sin\!\big(\Delta\theta_{1m}(k)-(m-1)\Delta\theta\big).
\end{split}
\label{eq:score_m}
\end{equation}
The second derivative is therefore
\begin{equation}
\begin{split}
\frac{\partial^2 \ell_m(\Delta\theta)}{\partial \Delta\theta^2}
&=
\kappa_m (m-1)\frac{\partial}{\partial \Delta\theta}
\sin\!\big(\Delta\theta_{1m}(k)-(m-1)\Delta\theta\big)
\\
&=
-\kappa_m (m-1)^2 \cos\!\big(\Delta\theta_{1m}(k)-(m-1)\Delta\theta\big).
\end{split}
\label{eq:hess_m}
\end{equation}
By definition, the Fisher information for $\Delta\theta$ contributed by baseline $m$ is
\begin{equation}
\begin{split}
\mathcal{I}_m(\Delta\theta)
&=
-\mathbb{E}\!\left[
\frac{\partial^2 \ell_m(\Delta\theta)}{\partial \Delta\theta^2}
\right] \\
&=
\kappa_m (m-1)^2\,
\mathbb{E}\!\left[
\cos\!\big(\Delta\theta_{1m}(k)-(m-1)\Delta\theta\big)
\right],
\end{split}
\label{eq:FI_m_step1}
\end{equation}
where the expectation is taken with respect to \eqref{eq:vonmises_phi}. Under a von Mises distribution with mean $\mu$ and concentration $\kappa$, a standard moment identity is
\begin{equation}
\mathbb{E}\!\left[\cos(\phi-\mu)\right]=\frac{I_1(\kappa)}{I_0(\kappa)},
\label{eq:vm_moment}
\end{equation}
where $I_1(\cdot)$ is the modified Bessel function of the first kind (order one). Applying \eqref{eq:vm_moment} with $\mu=(m-1)\Delta\theta$ yields
\begin{equation}
\mathcal{I}_m(\Delta\theta)
=
\kappa_m (m-1)^2\,
\frac{I_1(\kappa_m)}{I_0(\kappa_m)}.
\label{eq:FI_m}
\end{equation}
Assuming conditional independence across baselines, the total Fisher information is additive:
\begin{equation}
\begin{split}
\mathcal{I}_{\Delta\theta}(k)
&=
\sum_{m=2}^{M}\mathcal{I}_m(\Delta\theta)
=
\sum_{m=2}^{M}\kappa_m (m-1)^2 \,\rho(\kappa_m), \\
\rho(\kappa)
&\triangleq
\frac{I_1(\kappa)}{I_0(\kappa)}.
\end{split}
\label{eq:FI_delta}
\end{equation}
The scaling factor $\rho(\kappa) \in (0,1)$ represents the Fisher information efficiency relative to the linear Gaussian model. It exhibits the asymptotic behavior $\rho(\kappa) \to 1$ in the high-concentration regime ($\kappa \to \infty$), recovering the Euclidean bound, while decaying as $\rho(\kappa) \approx \kappa/2$ for vanishing concentration ($\kappa \to 0$).

The CRLB for any unbiased estimator of $\Delta\theta$ follows as
\begin{equation}
\begin{split}
\mathrm{Var}\!\big(\widehat{\Delta\theta}\big)
\;\ge\;
\frac{1}{\mathcal{I}_{\Delta\theta}}
&=
\left(
\sum_{m=2}^{M}\kappa_m (m-1)^2 \,\rho(\kappa_m)
\right)^{-1}.
\end{split}
\label{eq:CRLB_delta_exact}
\end{equation}
In the high-concentration regime, $\rho(\kappa_m)\approx 1$, yielding the simplified scaling law
\begin{equation}
\mathrm{Var}\!\big(\widehat{\Delta\theta}\big)
\;\gtrsim\;
\left(
\sum_{m=2}^{M}\kappa_m (m-1)^2
\right)^{-1}.
\label{eq:CRLB_delta}
\end{equation}

\noindent Finally, using $\Delta\theta_{\text{DoA}}=(2\pi d/\lambda)\sin\phi$ and the scalar CRLB transformation rule, we obtain
\begin{equation}
\begin{split}
\mathrm{Var}\!\big(\widehat{\phi}\big)
&\ge
\frac{1}{\left(\frac{\partial \Delta\theta_{\text{DoA}}}{\partial \theta}\right)^{\!2}\,\mathcal{I}_{\Delta\theta}}
\\
&=
\left[
\left(\frac{2\pi d}{\lambda}\cos\phi\right)^{\!2}
\sum_{m=2}^{M}\kappa_m (m-1)^2 \,\rho(\kappa_m)
\right]^{-1}.
\end{split}
\label{eq:CRLB_theta_exact}
\end{equation}
The bound in \eqref{eq:CRLB_theta_exact} explicitly reveals the aperture-driven gain through $\sum_{m=2}^{M}(m-1)^2$ and the classical degradation near endfire through $\cos\phi_k$.

\subsection{Localization Precision}
\label{subsec:localization}

The DoA bound in \eqref{eq:CRLB_theta_exact} translates directly into localization precision when the receiver knows the DAS node locations and the active-node schedule (or, equivalently, can decode the active index). Since the transmitter alternates among spatially separated radiators, the receiver collects bearing observations from multiple known anchor points over time, enabling bearing-only localization (triangulation) with quantifiable accuracy.

Consider 2D localization for clarity (the extension to 3D follows by using azimuth/elevation). Let the unknown device position be
\(
\mathbf{p}\triangleq [x\;\;y]^\top\in\mathbb{R}^2
\)
and let the $i$-th DAS radiator (anchor) be located at a known position
\(
\mathbf{p}_i\triangleq [x_i\;\;y_i]^\top.
\)
At time $k$, the active radiator index is $I_k\in\{1,\ldots,N_{\mathrm{tx}}\}$ and the dominant DoA observed at the receiver is $\phi_k$. Under line-of-bearing sensing, the noiseless bearing associated with anchor $i$ is
$\phi^{(i)}(\mathbf{p})
=
\atan2\!\big(y-y_i,\;x-x_i\big),$
and the corresponding observation model is
$\widehat{\phi}_k
=
\phi^{(I_k)}(\mathbf{p})
+
\omega_k.$
For small error, we approximate $\omega_k$ as zero-mean with variance lower bounded by \eqref{eq:CRLB_theta_exact}, and define
\begin{equation}
\begin{split}
\sigma_{\phi,k}^2
&\triangleq
\mathrm{Var}(\widehat{\phi}_k)
\\
&\gtrsim
\Bigg[
\left(\frac{2\pi d}{\lambda}\cos\phi_k\right)^{\!2}
\sum_{m=2}^{M}\kappa_m (m-1)^2 \,\rho(\kappa_m)
\Bigg]^{-1}.
\end{split}
\label{eq:sigma_theta_from_CRLB}
\end{equation}
Collect $K$ bearing measurements over time. A natural estimator of $\mathbf{p}$ is the Weighted Least Squares (WLS) problem
\begin{equation}
\begin{aligned}
\widehat{\mathbf{p}}
\in
\arg\min_{\mathbf{p}\in\mathbb{R}^2}
\sum_{k=1}^{K}
w_k\,\big|\wrap\!\big(\widehat{\phi}_k-\phi^{(I_k)}(\mathbf{p})\big)\big|^2,
\quad
w_k \triangleq \sigma_{\phi,k}^{-2},
\end{aligned}
\label{eq:WLS_localization}
\end{equation}
where $\wrap(\cdot)$ enforces a small-angle residual in $(-\pi,\pi]$. The weighting $w_k=\sigma_{\phi,k}^{-2}$ is information-optimal in the small-error regime and couples localization quality to the DoA accuracy through \eqref{eq:sigma_theta_from_CRLB}.

To expose a closed-form precision law, we linearize the measurement model $\widehat{\phi}_k = \phi^{(I_k)}(\mathbf{p}) + \omega_k$ about the true $\mathbf{p}$. For anchor $i$, define the relative displacement and range
\[
\Delta x_i \triangleq x-x_i,\quad
\Delta y_i \triangleq y-y_i,\quad
r_i^2 \triangleq \Delta x_i^2+\Delta y_i^2.
\]
Using the definition $\phi^{(i)}(\mathbf{p}) \triangleq \atan2(y-y_i, x-x_i)$, the gradient of the bearing function is derived as
\begin{equation}
\nabla_{\mathbf{p}}\,\phi^{(i)}(\mathbf{p})
=
\frac{1}{r_i^2}
\begin{bmatrix}
-\Delta y_i\\
\ \Delta x_i
\end{bmatrix}.
\label{eq:bearing_jacobian}
\end{equation}
Assuming (approximately) independent bearing errors across time, the (2D) Fisher information matrix (FIM) for $\mathbf{p}$ is
\begin{equation}
\mathbf{J}(\mathbf{p})
=
\sum_{k=1}^{K}
\frac{1}{\sigma_{\phi,k}^2}
\Big(\nabla_{\mathbf{p}}\,\phi^{(I_k)}(\mathbf{p})\Big)
\Big(\nabla_{\mathbf{p}}\,\phi^{(I_k)}(\mathbf{p})\Big)^\top.
\label{eq:FIM_position_general}
\end{equation}
Substituting \eqref{eq:bearing_jacobian} yields an explicit contribution from each active anchor:
\begin{equation}
\mathbf{J}(\mathbf{p})
=
\sum_{k=1}^{K}
\frac{1}{\sigma_{\phi,k}^2\,r_{I_k}^4}
\begin{bmatrix}
\Delta y_{I_k}^2 & -\Delta x_{I_k}\Delta y_{I_k}\\[2pt]
-\Delta x_{I_k}\Delta y_{I_k} & \Delta x_{I_k}^2
\end{bmatrix}.
\label{eq:FIM_position_explicit}
\end{equation}
The localization CRLB follows as
$\mathrm{Cov}(\widehat{\mathbf{p}})
\succeq
\mathbf{J}(\mathbf{p})^{-1},$ and a standard scalar summary is the \emph{position error bound} (PEB),
\begin{equation}
\mathrm{PEB}
\triangleq
\sqrt{\mathrm{tr}\!\left(\mathbf{J}(\mathbf{p})^{-1}\right)},
\label{eq:PEB}
\end{equation}
which quantifies the best achievable Root Mean Squared Error (RMSE) under unbiased estimation.

Expressions in \eqref{eq:FIM_position_general}--\eqref{eq:PEB} make the dependence of localization precision on the DoA accuracy explicit via $\sigma_{\phi,k}^2$. Specifically, improving manifold-domain sensing tightens the bound through \eqref{eq:sigma_theta_from_CRLB} and \eqref{eq:FIM_position_general}; larger effective receive aperture increases $\sum_{m=2}^{M}\kappa_m (m-1)^2\rho(\kappa_m)$ and reduces $\sigma_{\phi,k}^2$, while unfavorable anchor geometry (nearly collinear bearings) leads to an ill-conditioned $\mathbf{J}(\mathbf{p})$ and a looser PEB. Moreover, \eqref{eq:FIM_position_explicit} shows that each anchor contribution scales as $r_i^{-4}$, reflecting reduced bearing sensitivity with range and motivating anchor placement and scheduling that yield diverse bearings at sufficiently small ranges whenever possible.

\section{Simulation Results}
\label{sec:sim}

We evaluate the performance of the proposed SPMC in terms of communication and DoA-sensing capabilities. Even though the proposed approach and receiver architecture are LO-free, we benchmark their performance against an LO-based coherent receiver employing carrier/phase recovery followed by coherent I/Q processing. Unless stated otherwise, we consider the following setting. Communications are carried out over a carrier frequency of \(f_c = 28\) GHz. The transmitter is equipped with \(N_{\mathrm{tx}} = 16\) spatially separated antennas and operates over a Rician fading channel, which is equivalent to employing a 16-order modulation scheme. The phase noise, arising from transmitter imperfections such as LO imperfection and time jitter, is assumed to be unknown and to cause a random phase shift with zero mean and standard deviation \(\sigma_{\phi} = 10^\circ\). This phase noise is applied in addition to the thermal noise.

\subsection{Communication Performance}
\label{subsec:sim_comm}

\begin{figure}[h]
    \centering
    \includegraphics[width=\linewidth]{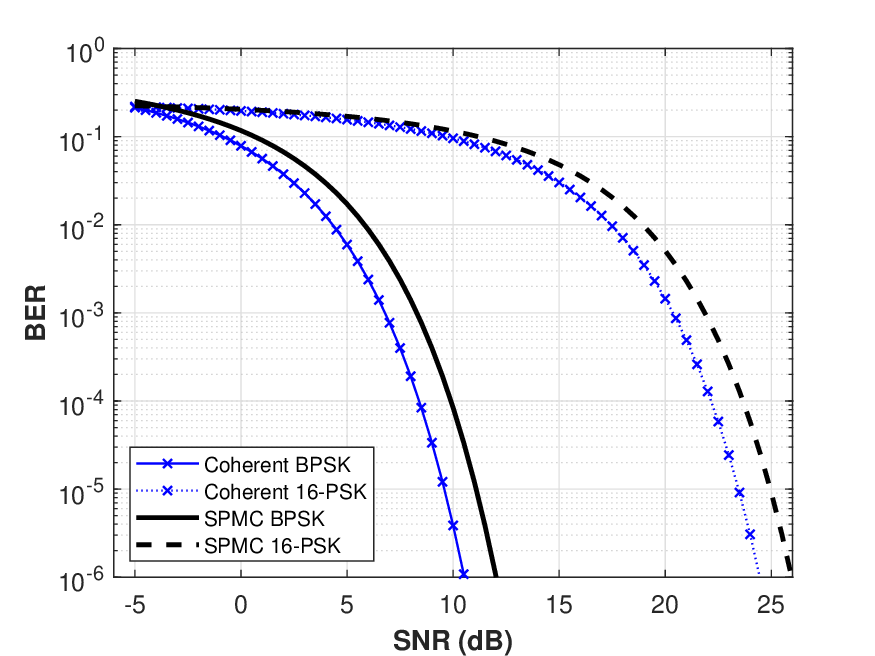}
    \caption{BER in the absence of phase noise. }
    \label{fig:bpsk_16psk_ber_Wo_PN}
    \vspace{-0.3cm}
\end{figure}

\begin{figure}[h]
    \centering
    \includegraphics[width=\linewidth]{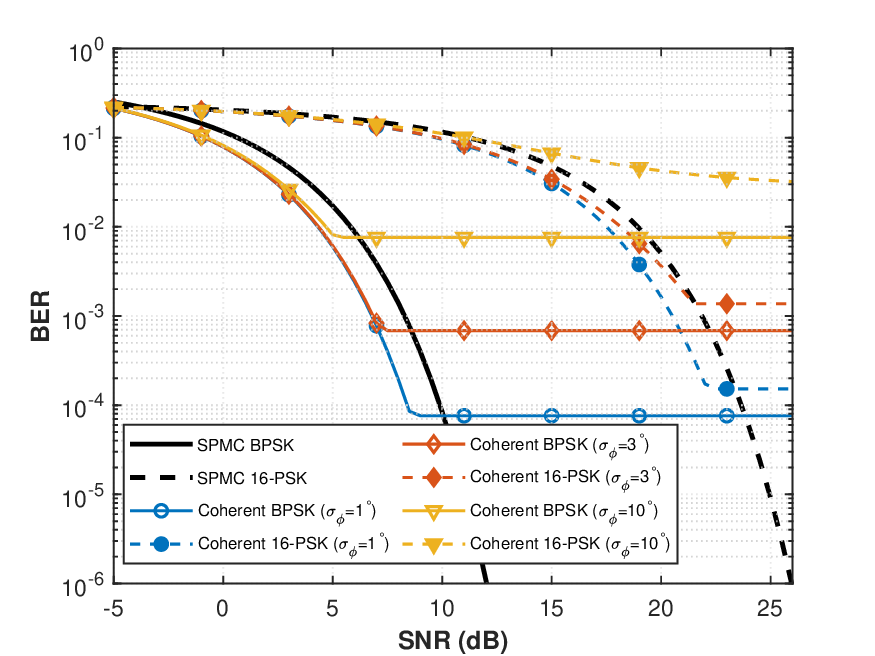}
    \caption{BER performance as a function of SNR for different phase-noise variances $\sigma_{\phi}\in\{1^{\circ},3^{\circ},10^{\circ}\}$}
    \label{fig:bpsk_16psk_ber_W_PN}
    \vspace{-0.3cm}
\end{figure}

The proposed technique offers two practical benefits: operation without a LO for coherent carrier/phase recovery, and invariance to phase noise perturbations introduced by LO imperfection. For completeness, it remains informative to benchmark the bit-error-rate (BER) performance against a coherent (LO-based) receiver under idealized conditions with negligible phase noise, and then to quantify robustness as phase noise increases. Figures~\ref{fig:bpsk_16psk_ber_Wo_PN} and~\ref{fig:bpsk_16psk_ber_W_PN} report the BER results in the absence and presence of phase noise, respectively. In Fig.~\ref{fig:bpsk_16psk_ber_Wo_PN} (no phase noise), the SPMC receiver exhibits a modest SNR penalty relative to coherent detection, while preserving the same qualitative BER trend; the gap remains within approximately $1$~dB over the operating region of interest. In Fig.~\ref{fig:bpsk_16psk_ber_W_PN} (with phase noise), the coherent receiver degrades markedly as $\sigma_{\phi}$ increases, including the emergence of error floors at high SNR, consistent with residual carrier/phase estimation errors. In contrast, SPMC depends on inter-antenna relative phase and is therefore insensitive to phase noise; accordingly, BER continues to decrease monotonically with SNR and no error floor is observed over the simulated parameter range.

\subsection{Sensing Performance}
\label{subsec:sim_sensing}

\begin{figure}[t]
    \centering
    \includegraphics[width=\linewidth]{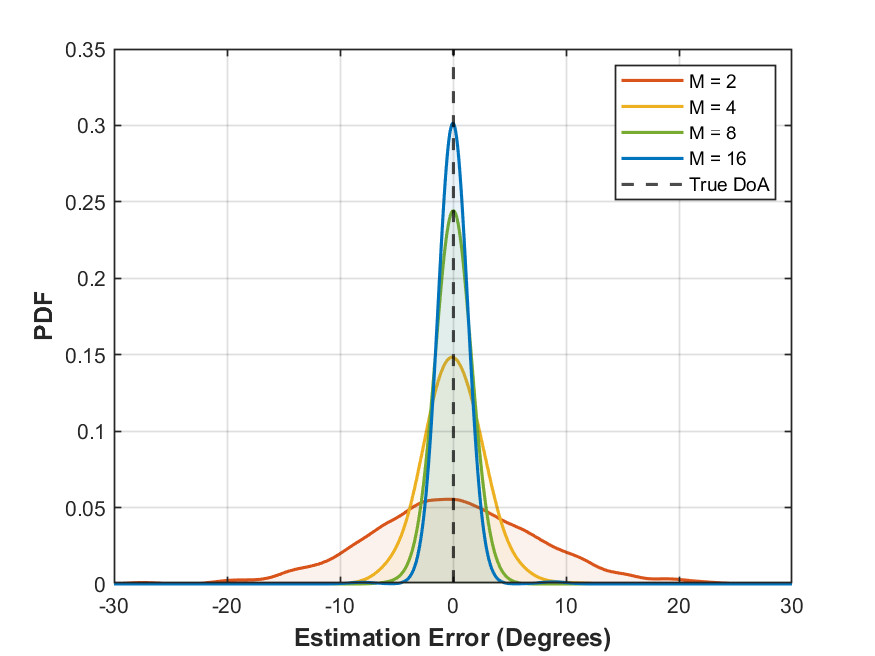}
  \caption{DoA estimation error distribution for varying values of $M$. }
    \label{fig:spatial_precision_pdf_alt}
    \vspace{-0.3cm}
\end{figure}

DoA-sensing performance is evaluated using the correlation-domain observables produced by the SPMC front-end, i.e., the same base-band statistics used for data extraction are also used as the input to the DoA inference stage. Fig.~\ref{fig:spatial_precision_pdf_alt} reports the empirical probability density function (PDF) of the DoA estimation error for different array sizes (M). As the number of antenna elements increases from (M=2) to (M=16), the error PDF becomes progressively narrower and more concentrated around zero. This concentration indicates a reduction in the estimator dispersion (lower variance) and a lower probability of large outliers, consistent with the increased effective aperture and improved spatial resolution afforded by larger arrays.

Overall, the systematic narrowing of the error PDFs in Fig.~\ref{fig:spatial_precision_pdf_alt} demonstrates that increasing (M) improves DoA estimation accuracy when using SPMC correlation-domain measurements. The results further indicate that this accuracy improvement is retained under oscillator phase noise impairment, since the sensing pipeline operates on inter-antenna relative phase/correlation statistics rather than on LO-dependent coherent baseband samples.

\begin{figure}[t]
    \centering
    \includegraphics[width=\linewidth]{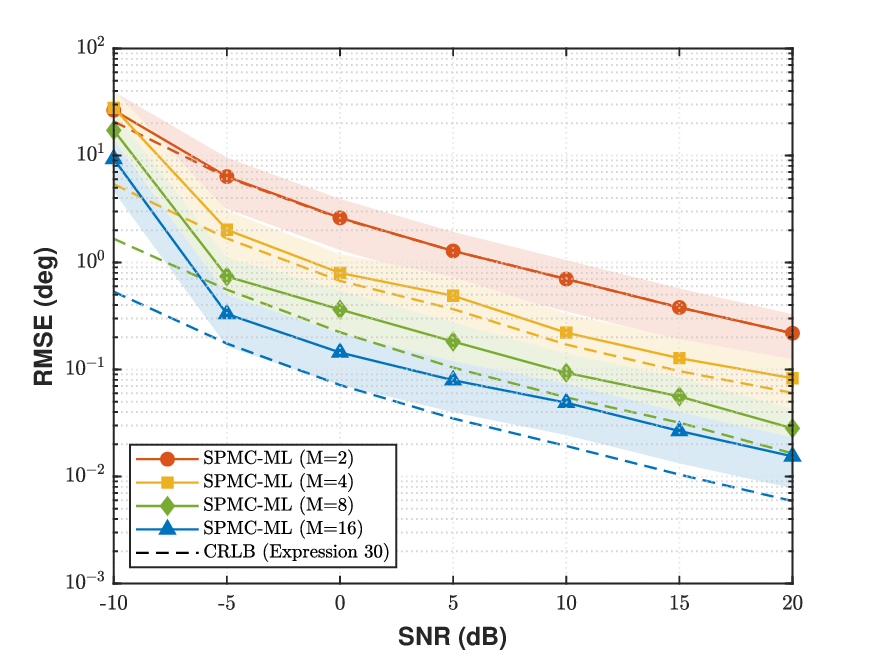}
    \caption{DoA sensing accuracy under phase noise: RMSE of the proposed SPMC--ML estimator versus SNR for different numbers of receive antennas $M$, compared with the CRLB in \eqref{eq:CRLB_theta_exact}.}
    
    \label{fig:CRLB}
    \vspace{-0.3cm}
\end{figure}

Fig.~\ref{fig:CRLB} quantifies the DoA-sensing performance obtained from the correlation-domain observables of the proposed SPMC front-end under phase perturbations. Across all array sizes, the DoA RMSE decays monotonically with SNR, indicating that the estimator remains stable and does not exhibit error-floor behavior over the simulated range. Increasing the number of active receive antennas from $M=2$ to $M=16$ yields a consistent reduction in RMSE, which is in line with the increased effective aperture and the corresponding Fisher-information scaling captured by \eqref{eq:FI_delta}--\eqref{eq:CRLB_theta_exact}. Furthermore, the simulated SPMC--ML curves follow the analytical CRLB trend in \eqref{eq:CRLB_theta_exact}, particularly in the moderate-to-high SNR regime, thereby providing numerical evidence that the proposed manifold-domain model and multi-baseline fusion accurately characterize the sensing precision achievable without LO-based carrier/phase tracking.

\section{Conclusion}
\label{sec:conclusion}

We propose SPMC, an LO-free receiver JCAS framework that embeds information in the inter-antenna phase and recovers it via antenna-domain correlation without an LO and without I/Q coherent demodulation. By projecting correlator outputs onto the unit-circle manifold, SPMC achieves amplitude invariance and enables native joint communication-and-sensing through operator-based decomposition (inter-antenna phase difference for data recovery and DoA estimation). We develop a quadrature correlator receiver, provide identifiability conditions, and derive representative error bounds under wrapped phase noise. The results are promising and demonstrate the potential of SPMC as a solution for integrated sensing and communication for IoT. Future work includes hardware prototypes at mmWave/sub-THz.

\section*{Acknowledgment}
This work was supported by The Scientific and Technological Research Council of Türkiye (TÜBİTAK) under project number 125E370.


\end{document}